\documentstyle[aps,epsf,floats, manuscript]{revtex}
\begin{document}

\title{Mirror symmetry breaking through an internal degree of freedom leading
  to directional motion} \author{S. Cilla\footnote{To whom correspondence
    should be addresed: scilla@posta.unizar.es}, F. Falo, L.M. Flor\'{\i}a}
\address{ Departamento de F\'{\i}sica de la Materia Condensada. Departamento
  de Teor\'{\i}a y Simulaci\'on de Sistemas Complejos. Instituto de Ciencia de
  Materiales de Arag\'on CSIC-Universidad de Zaragoza, 50009 Zaragoza, Spain}

\date{\today} \maketitle

\begin{abstract}
  We analyze here the minimal conditions for directional motion (net flow in
  phase space) of a molecular motor placed on a mirror-symmetric environment
  and driven by a center-symmetric and time-periodic force field. The complete
  characterization of the deterministic limit of the dissipative dynamics of
  several realizations of this minimal model, reveals a complex structure in
  the phase diagram in parameter space, with intertwined regions of pinning
  (closed orbits) and directional motion. This demonstrates that the
  mirror-symmetry breaking which is needed for directional motion to occur,
  can operate through an internal degree of freedom coupled to the
  translational one.
\end{abstract}

\pacs{PACS numbers: 05.40$-$a,05.45$-$a,87.10$+$e}

The possibility of motion rectification at brownian scale from non equilibrium
fluctuations is an interesting question already posed by Smoluchowski
\cite{Smo} and Feynman \cite{Feynman}. It has been shown that directional
motion (appearance of net flow in phase space) of brownian particles, without
any macroscopic gradient or force, can be achieved provided that the potential
exhibits spatial asymmetry and detailed balance is broken. This latter
condition assures that the system is out of equilibrium. Systems with a
periodic potential profile but spatial asymmetry, called \emph{ratchets}, have
been addressed as systems in which non equilibrium fluctuations can induce
directional motion \cite{Magnasco}. They have attracted much attention on the
basis that they can help to understand the physics of molecular motors
\cite{Julicher}, along with the possibilities they open for using these ideas
in superconductors \cite{Lee}, Josephson junctions \cite{Zapata}, quantum dots
\cite{Linke} and in the
promising world of nanotechnology \cite{Rouse}.

Hereon, we will focus on "motor" systems such that their internal degrees of
freedom are essential for net directional motion to occur. The motivation for
this problem came initially from the (bio)molecular motors field when it was
found \cite{Henni} that kinesin direction of motion along microtubules could
be reversed by modifying the architecture of a small domain of the protein
called "neck region". These discoveries could suggest that the mirror
symmetry-breaking mechanism responsible for the directional motion performed
by these proteins could lie in their own structure rather than in their
environment. In this paper we will consider the type of  systems
 with a mirror-symmetric
environment for its traslational degree of freedom ($u^{tr}$). The
symmetry-breaking mechanism acts through an internal degree of freedom
($u^{int}$) coupled to it, that is, we will consider a dimer model (two
degrees of freedom). We are interested in the minimal conditions for the
operation of this system as a motor (being capable of moving against an
applied field). Previous works have also analysed directional motion in
dimer models \cite{finite} but the potential for $u^{tr}$
 considered was ratchet \emph{ab initio}. On the contrary, the dimer model we present in this paper is
inmersed in a symmetric environment. This is also the case  for the system
recently studied by \cite{Porto}. There, internal degrees of freedom (more
than two) experience a flashing interaction potential, while ours is a rocked
system. After a complete characterization of the
deterministic limit of the dynamics, which reveals the basic nonlinear
mechanics of directional motion, we will briefly 
discuss the gross features of the Langevin dynamics of this
model and its utility as a help to design new technologies at micro and
nanoscales.

As a working mechanical visualization of this model, one can use two
overdamped and coupled brownian particles, with positions $u_{1}$ and $u_{2}$
(so $u^{tr}=1/2(u_{1} +u_{2})$ and $u^{int}=u_{2}-u_{1}$) moving in a
periodic, symmetric potential $V(u+1)=V(u)=V(-u)$ and being driven by
center-symmetric periodic forces $F_{i}=-F_{i}(t+T/2)\hspace{0.5mm} i=1,2$ of
period $T=2\pi/w$ \cite{Note}. In the deterministic limit, the equations of
motion read
\begin{equation}
\dot{u}_{1}=-V'(u_{1}) -\partial_{1}W(u_{1},u_{2}) + F_{1}(t) 
\end{equation}
\begin{equation}
\dot{u}_{2}=-V'(u_{2}) -\partial_{2}W(u_{1},u_{2}) + F_{2}(t) 
\end{equation}     
We impose on $W(u_{1},u_{2})$ the general condition of being a function of the
relative distance $u^{int}=u_{2}-u_{1}$, so the partial derivatives with respect to
$u_{1}$ and $u_{2}$ verify $\partial_{1}
W(u_{1},u_{2})=-\partial_{2}W(u_{1},u_{2})$. We  will consider the cases in
which $W$ is a convex and a non-convex function of $u^{int}$.

Equations (1)-(2) remain invariant under the symmetry transformations
$(u_{1},u_{2},t)\rightarrow(-u_{2},-u_{1},t + T/2)$ provided that
$F_{1}(t)=F_{2}(t)$ and $(u_{1},u_{2})\rightarrow(-u_{2},-u_{1})$ if
$F_{1}(t)=-F_{2}(t)$. With this proviso one can easily show that any averaged
"directional motion" in phase space is necessarily (and
straightforwardly) null : if there exists a
solution of (1)-(2) with nonzero velocity, by symmetry we can find another
solution with the same velocity but opposite sign, so no net motion can be
observed when averaging over the phase space. Directional motion in this
strong sense can only occur if the inequality $F_{1}(t)\neq \pm F_{2}(t)$
holds, regardless the specific form (convex or non-convex) of the
interaction potential $W$. Taking $F_{i}(t)$ from the class of functions
$F(t)=F_{ac}\sin(\omega t+2\pi\delta)$ (the simplest periodic center-symmetric
function  $F(t)=-F(t+ T/2)$), there are two ways of breaking the
symmetry of this system ($F_{1}(t)\neq \pm F_{2}(t)$):
\begin{itemize}
\item \emph{i)} $max_{t} F_{1}(t)\neq max_{t} F_{2}(t)$
\item \emph{ii)} $\delta_{1}\neq \delta_{2} \bmod \frac{1}{2}$ 
\end{itemize}
That is, applying forces on each particle with (\emph{i}) different amplitude
 or (\emph{ii}) different phase. We will focus on the limit cases
for each situation: we will consider $F_{ac}^{(1)}(t)=0$ and
$F_{ac}^{(2)}(t)\neq 0$ as well as $\delta_{1}=0$ and $\delta_{2}\neq 0$. Any
other situation verifying the condition $F_{1}(t)\neq \pm F_{2}(t)$ is just a
  combination of these
two possibilities. 

In particular, most numerics have been performed with the simplest choices
\begin{equation}
V(u)=\frac{Q}{(2\pi)^{2}}\cos(2\pi u) \hspace{0.5cm}
W(u_{1},u_{2})=\frac{K}{2}[(u_{2}-u_{1}) -l_{0}]^{2}
\end{equation}
The convexity of $W$ and the dissipative character of the dynamics have the
important consequence that the partial order among initial conditions is
preserved (monotonicity property or "no-passing rule" \cite{Middleton}). The
monotonicity property says that if at a time $t=t_{0}$ two initial conditions
${\bf u}$,${\bf v}$ verify that $u_{1}(t_{0})<v_{1}(t_{0})$ and
$u_{2}(t_{0})<v_{2}(t_{0})$, then for any time $t>t_{0}$ the inequalities
$u_{1}(t)<v_{1}(t)$ and $u_{2}(t)<v_{2}(t)$ hold. This property 
leaves small room for
deterministic complexities of this non-integrable dynamics. For example, the
asymptotic mean velocity of all trajectories in the phase space is unique, and
vibrating pinned solutions (closed orbits) cannot coexist with mobile ones.
The choice of $W(u_{1},u_{2})$ implies another symmetry relation: the system
remains invariant under the transformation $(u_{1}, u_{2}, l_{0})\rightarrow
(-u_{1}, -u_{2}, 1-l_{0})$. That is, if we have a mobile solution of (1)-(2)
for a value $l_{0}$ we have another mobile solution with opposite velocity for
$1-l_{0}$. For fixed values of $(Q,K,l_{0})$ we have approximated
numerically to optimal accuracy the function ${\mathcal J}(F_{ac},\omega)$
where ${\mathcal J}$ is defined as the asymptotic flow
in phase space ${\mathcal J}=\langle\dot{u}^{tr}\rangle$. 

In figures 1.a) and 1.b) the ${\mathcal J}(F_{ac})$ profiles at typically low
$(\omega=2\pi\times 0.01)$ and high frequencies $(\omega=2\pi\times 0.1)$ for
$F_{1}(t)=0$ and $F_{2}(t)=F_{ac}\sin(\omega t)$ (case \emph{i)} ---different
amplitudes---) are shown for the values $(Q,K,l_{0})=(2,1,1/4)$. As can be
seen from  figure 1.a, the complex stairlike structure of the relation
${\mathcal J}(F_{ac})$ for low frequencies is qualitatively indistinguishable
from those exhibited by single particle rocking ratchets (asymmetric potential).  However, when the frequency is
increased, the direction of motion for fixed $l_{0}$ 
can be either positive or negative (figure
1.b), something which cannot occur in deterministic single particle
rocking ratchets. The primary structure of mobility bands in $F_{ac}$ 
separated by
vibrating pinned solutions is well understood in terms of saddle-node
bifurcations signaling the onset of global flow in phase space. Figure 1.c
shows the bifurcation diagram of pinned (vibrating) solutions. $\phi$
represents the mean value over a period of the external force of the
traslational variable, $\phi=\langle u^{tr}\rangle_{T}$. Thick lines represent
stable orbits whereas  thin ones represent unstable ones. This bifurcation
diagram  neatly  determines the intervals of motion in $F_{ac}$ as well as the local sign of
${\mathcal J}$ in both interval's edges. The numerical inspection of the scaling of
intermittencies density is the one of a type I intermittency scenario
(associated here to quasiperiodicity and frequency locking), giving the
staircase aspect of ${\mathcal J}(F_{ac})$ close to the mobility onset.

In figure 2.c an example of ${\mathcal J}(F_{ac})$ profile for
$F_{1}(t)=F_{ac}\sin(\omega t)$, $F_{2}(t)=F_{ac}\sin(\omega t +2\pi\delta)$
(case \emph{ii)} ---different phase---) with
$(\omega,\delta)=(2\pi\times0.05,0.35)$, $(Q,K,l_{0}=(2,1/2,1/4)$ shows clearly the complex alternation of positive and
negative flow in the phase space.  Again the mobility bands in $F_{ac}$
when $(\omega, \delta)$ remain fixed, appear after the collision of stable and
unstable orbits by means of a saddle-node bifurcation. 
In figure 2.a) we have plotted for $\omega=2\pi\times 0.05$ the width in
$F_{ac}$ of the mobility bands at different values of $\delta$. Figure 2.b)
shows the values of $F_{ac}$ at which the first and second bifurcations occur,
that is, the smaller value of $F_{ac}$ at which the stable and unstable orbits
collide (depinning transition) and the nearest value at which they emerge again
(pinning transition) what determines the width of the first mobility band. 
As usual, very simple dynamics reveal
astonishingly complex phase diagrams whenever time and length scale
competition plays a role.

We want to stress that the rectification mechanisms in the models presented
here are completely different from other rocking ratchet systems with internal
degrees of freedom (d.o.f) \cite{finite}, where the symmetry is broken
\emph{ab initio} by the asymmetric potential $V(u)$. The only system with a
symmetric environment for the traslational d.o.f. in which symmetry breaking
comes through the internal d.o.f. is the flashing system (as oposed to
rocking) studied by Porto et al. \cite{Porto}. We also remark that the
rectification mechanisms for the cases numerically studied in the preceeding
paragraphs show some differences: in case (\emph{i}) we observe directional
motion even at the adiabatic limit (slow varying forces) whereas in case
(\emph{ii}) directional motion occurs only at finite frequencies. When
applying forces with different amplitudes, the ratchet effect lies in the fact
that one particle acts as a cargo, so it is easier to move the system in the
"driver" particle direction than in the other. In figure 3 it can be clearly
seen that the depinning force is smaller in the driver's direction than in the
other.  When applying different phases, the ratchet effect is more subtle: the
combined effect of phase value and strength $F_{ac}$ determines which particle
plays the role of cargo and which one the role of driver. The absence of the
adiabatic limit for this case (\emph{ii}) precludes the use of time
independent schemes for understanding rectification in an intuitive way. 
In figure 4 two trajectories corresponding to both limit situations (cases
\emph{i)}, \emph{ii)} above) analyzed are drawn.  Although our system is a
rocking ratchet, there is an alternative view in which, looking at the
equations of motion for the traslational variable $u^{tr}$
\begin{equation}
\dot{u}^{tr}=\frac{Q}{2\pi}\sin(2\pi u^{tr})\cos(\pi u^{int})+\frac{1}{2}(F_{1}(t) +F_{2}(t))
\label{eq:fash}
\end{equation}
one can see for both situations, case \emph{i}) and \emph{ii}), that the time dependence (periodic or quasiperiodic) of
$u^{int}$ allows the interpretation of the first term in rhs as a flashing
potential  for $u^{tr}$.  The directional motion can thus be seen as the
result of the  adequate synchrony between
external force on $u^{tr}$ and the periodic flashing potential.

When the convexity condition  on the interaction potential $W(u_{1},u_{2})$ is
 removed and non-convex interaction (such as double-well or Lennard-Jones)
 potentials are considered, the monotonicity property (which severely
 restricts the complexity of the dynamics) is lost, and generically the space
 is partitioned in basins of attraction corresponding to attractors with
 different asymptotic velocities. The description of the dynamics  becomes
 thus more complex. The phase portrait shows, for certain regions of parameter
 values chaotic attractors and there appear new bifurcations (other than the
 observed for the convex case: pitchfork, generic saddle-node and stability 
interchange) as the parameters change. Anyway the existence of net flow
 (ie. non zero phase space average of asymptotic velocities) in phase space is
 a generic property in wide regions of parameter space, as in the convex case.

We are dealing with rocking-like systems \cite{Magnasco} where the
rectification mechanism is completely deterministic, so when introducing noise
$\xi_{1}(t)$ and $\xi_{2}(t)$ in equations (1)-(2) (white gaussian noise with
correlation function
$\langle\xi_{i}(t)\xi_{j}(s)\rangle=D\delta_{i,j}\delta(t-s)$, $i,j=1,2$, being
$D=k_{B}T$ the diffusion coefficient) the mobility bands widen (fluctuation-induced depinning) and $\mathcal{J}$ decreases with increasing noise strength
(within the deterministic mobility bands). Both phenomena are easily explained
in the limit of small noise \cite{tesis}. For high $D$ the phenomenon of current
reversal\cite{currentrev} arises too and when noise strength is high 
enough diffusive motion dominates and no directional motion can be observed.

The proteins of the kinesin superfamily are just composed of two globular
"\emph{heads}" (catalitic domains that move over the microtubule using the 
energy delivered in ATP-hydrolysis). As mentioned before, recent experimental
 work suggests that the mechanism for directionality may rest on the
characteristics of the flexible structure joining both domains.
 ATP-hydrolysis
induces conformational changes in these proteins, that is, acts through
their internal degrees of freedom what may provide the symmetry-breaking
needed for the directional motion along the microtubule. The detailed
modelling of these conformational changes of the kinesin is beyond the scope
of this work; nevertheless the  generic model discussed in this paper could
help to understand  in the simplest mechanical terms the
role of the internal degrees of freedom in molecular motors.

Rigorous results on general simple models like this, could serve to guide or
inspire new technological applications specially in the new field of
nanotechnology. A simple nano-scale realization of the type of motors
considered here would consist of clusters of entities (particles or
macromolecular agregates) with different electrophoretic mobilities joined
with the aid of "flexible" molecules or polymers.  In order to reproduce the
symmetric environment it will suffice to place this engine in a row of
symmetric obstacles (electrodes for instance). When applying an ac electric
 field, because of the difference in the electrophoretic mobilities, 
different periodic forces will act on each cluster, making it possible to
 observe directional motion. \\
In summary, the analysis of these minimal models, convincingly demonstrate that
the mirror symmetry breaking needed for directional motion to occur, can act
through an internal degree of freedom eventhough the overall position of
the system experiences a symmetric environment.
\begin{center}

\end{center}
We want to acknowledge P.J. Mart\'{\i}nez for technical support and
 useful discussions, J.J. Mazo for his careful reading and suggestions. This
 work has  been financially supported by DGES (Spain) through the project
 PB98-1592. S. C. aknowledges a MEC  research grant.

\newpage
\begin{figure}[http]
\epsfxsize=2.5in
\centering{\mbox{\epsfbox{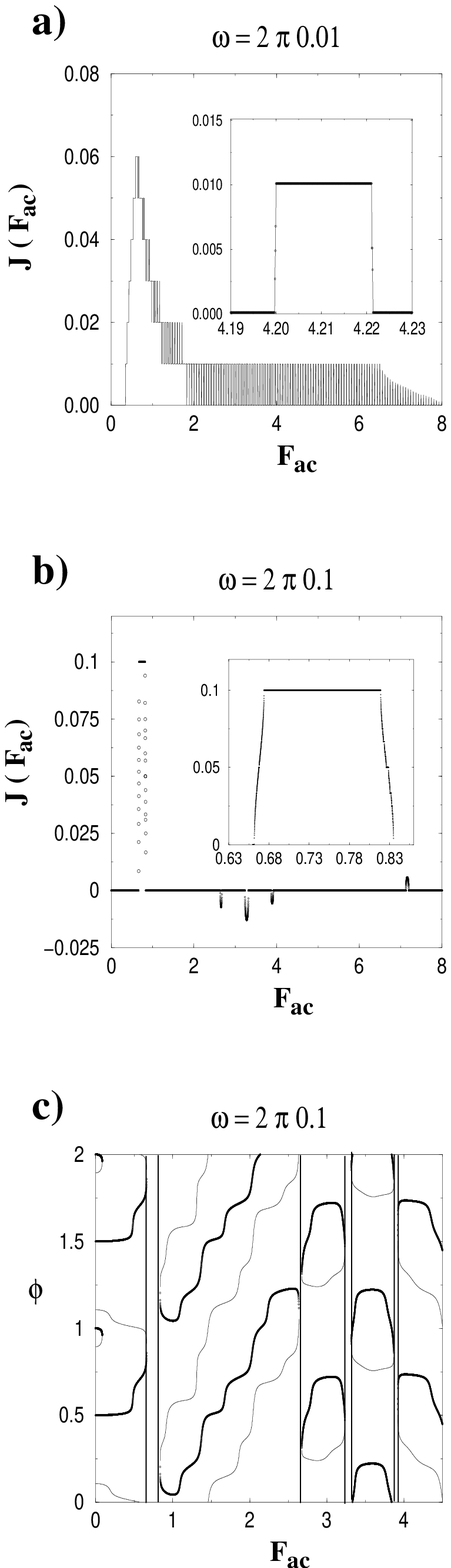}}}
\vspace{0.1in}
\caption[]{a) and b) Flow as a function of the amplitude of the external
  force $F_{ac}$  at two different frequency values. c) Bifurcation diagram
  using the mean value over a period of $ u^{tr}$ $\phi=\langle u^{tr}\rangle_{T}$ as relevant magnitude for $\omega=2\pi\times 0.1$. Intervals between vertical lines correspond to
  mobility bands in $F_{ac}$.}
\label{fig1}
\end{figure}
\newpage
\begin{figure}[http]
\epsfxsize=2.5in
\centering{\mbox{\epsfbox{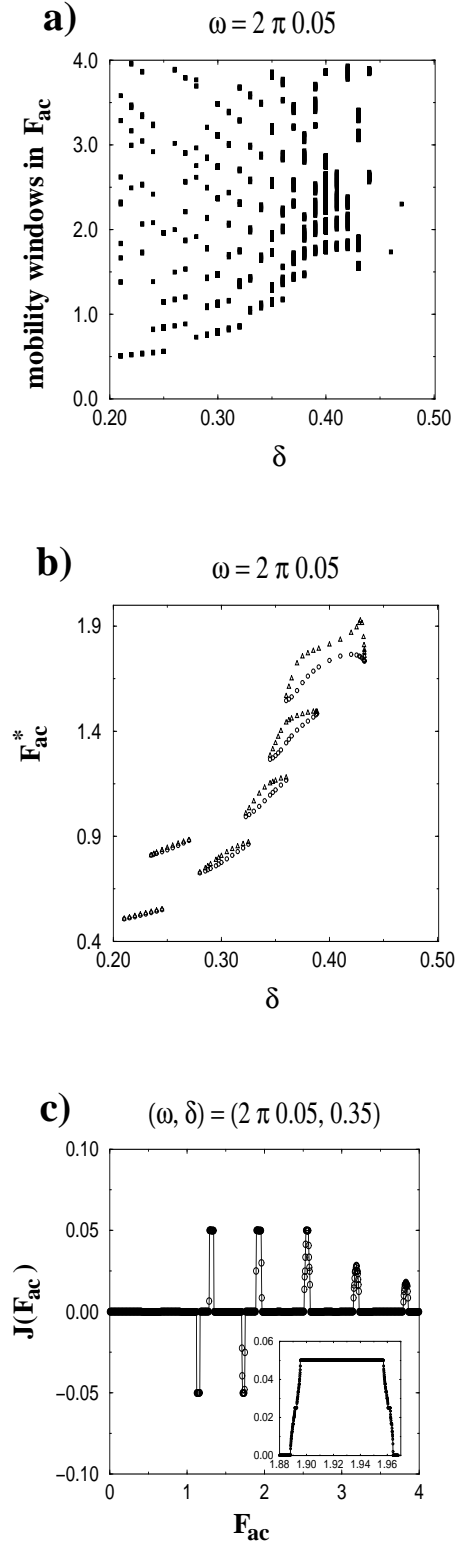}}}
\vspace{0.1in}
\caption{a) Mobility bands in $F_{ac}$ at $\omega=2\pi\times0.05$ as
  function of the phase difference $\delta$. b) Lower and upper values of
  $F_{ac}$ for the first mobility band as a function of $\delta$. c)
  Dependence of the velocity with $F_{ac}$ for $(\omega,\delta)=(2\pi\times
  0.05,0.35)$.}
\label{fig2}
\end{figure}
\newpage
\begin{figure}[http]
\epsfxsize=5.4in
\centering{\mbox{\epsfbox{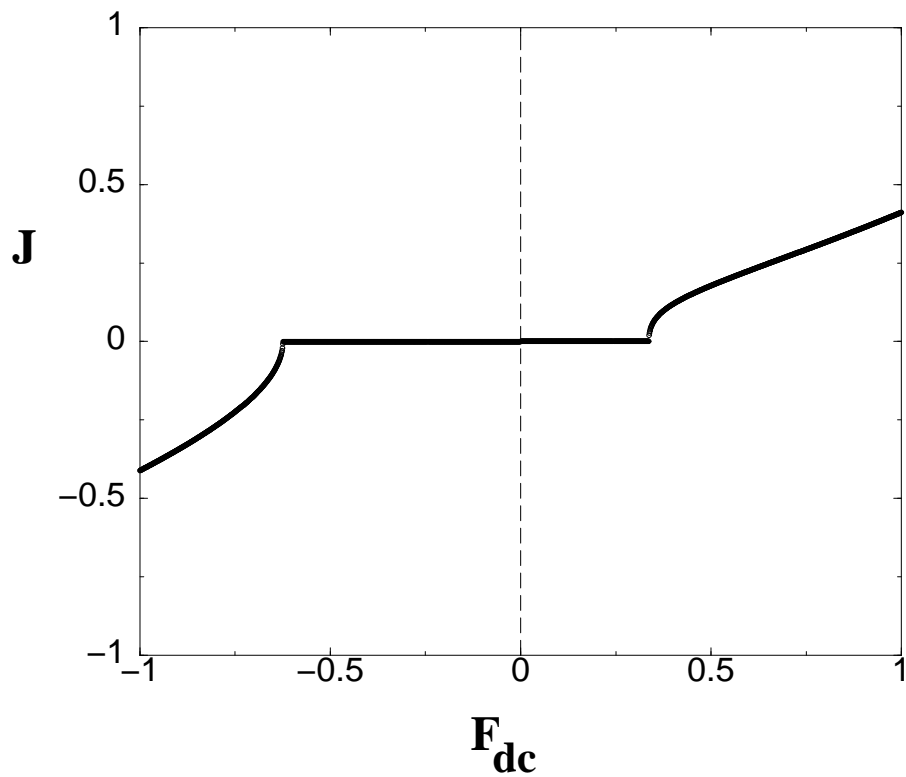}}}
\vspace{0.1in}
\caption{Flux ${\mathcal J}$ in the presence of a constant force: it can be
  clearly seen that the depinning force is smaller in the "driver" particle
 direction (positive one).}
\label{fig3}
\end{figure}

\newpage
\begin{figure}[http]
\epsfxsize=5.4in

\centering{\mbox{\epsfbox{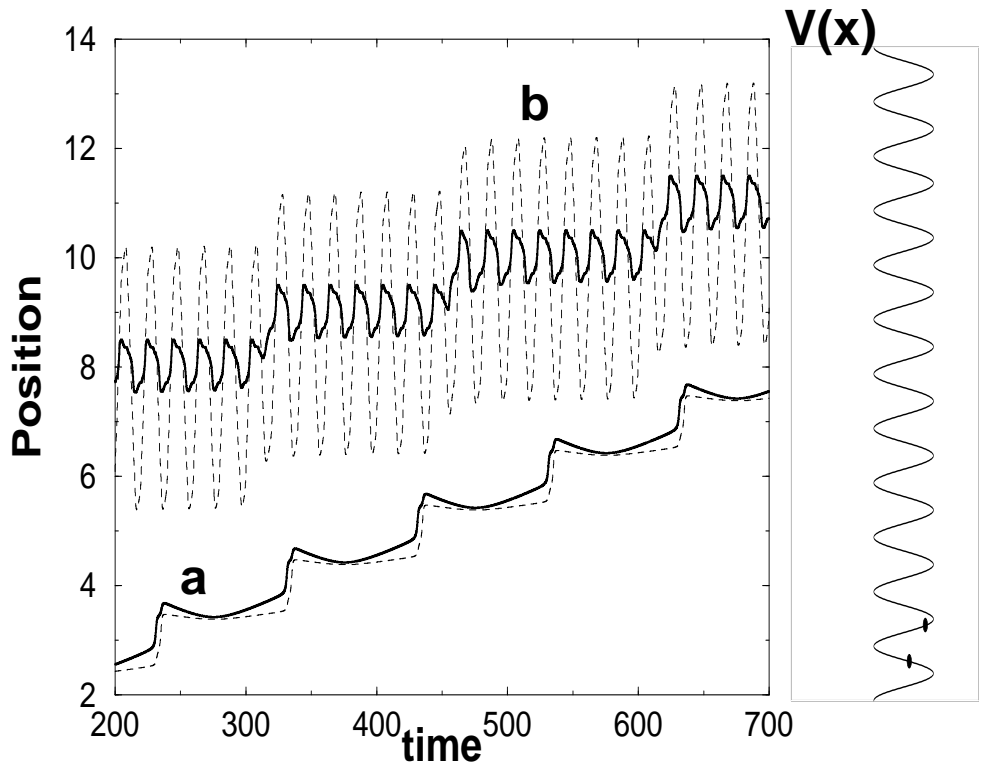}}}
\vspace{0.1in}
\caption{Deterministic trajectories ($u_{1}(t)$ dashed line and $u_{2}(t)$
  solid line)
  for the two models discussed. a) corresponds to different amplitudes (case
  \emph{i}) 
  $(\omega,F_{ac})=(2\pi\times0.01,0.35)$ and b to different phases (case
  \emph{ii}) $(\omega,F_{ac},\delta)=(2\pi\times0.05,1.284,0.35)$. }
\label{fig4}
\end{figure}

\end{document}